# RIP Twitter API: A Eulogy to Its Vast Research Contributions

Ryan Murtfeldt, Sejin Paik, Naomi Alterman, Ihsan Kahveci, Jevin D. West

## Abstract

Since 2006, Twitter's APIs have been rich sources of data for researchers studying social phenomena such as misinformation, public communication, crisis response, and political behavior. However, in 2023, Twitter began heavily restricting data access, dismantling its academic access program, and setting the Enterprise API price at $42,000 per month. Lacking funds to pay this fee, academics are scrambling to continue their research. This study systematically tabulates the number of studies, citations, publication dates, disciplines, and major topics of research using Twitter data between 2006 and 2024. While we cannot know exactly what will be lost now that Twitter data is cost-prohibitive, we can illustrate its research value during the years it was available. A search of eight databases found that between 2006 and 2024, a total of 33,306 studies were published in 8,914 venues, with 610,738 citations across 16 disciplines. Major disciplines include social science, engineering, data science, and public health. Major topics include information dissemination, tweet credibility, research methodologies, event detection, and human behavior. Twitter-based studies increased by a median of 25% annually from 2006 to 2023, but following Twitter's decision to charge for data, the number of studies dropped by 13%. Much of the 2024 research likely used data collected before the API shutdown, suggesting further decline ahead. This trend highlights a growing loss of empirical insight and access to real-time, public communication—raising concerns about the long-term consequences for studying society, technology, and global events in an era increasingly connected by social media.

## Introduction/Literature Review

### The Emergence of Social Media and the Rise of Platform Data

Commercial social media platforms burst to life between 2003-2006 with the launching of Myspace, Facebook, and Twitter (now called X, however for clarity we will refer to the platform as "Twitter" for the remainder of this paper) (Goodings, 2012; Lapowsky, 2019; Rothman, 2016). They quickly became a popular way for people to create and share content related to business, entertainment, science, crisis management, and politics (Stieglitz et al., 2018), with traffic growing by 24% between 2009 and 2010, accounting for 22% of the time people spend on the internet (The Nielsen Company, 2010). By 2009, 27% of online adults visited a social media site on any given day, and 73% of online American teens were using social media (Pew Research Center, 2010). By 2012, Facebook had nearly one billion users, and in 2013 was growing by 67% annually (Fan & Gordon, 2014).

## The Golden Age of Open, API Research

This growth of information sharing resulted in an enormous collection of data known as Social Media Big Data (Stieglitz et al., 2018; Tufekci, 2014), which included user's background information along with their daily activities on the platform such as likes and shares (Ghani et al., 2019). Shortly after launching their platforms, Twitter and Facebook opened their Application Programming Interfaces (APIs) in 2006 (Hylbert & Cosenza, 2020; Morin, 2008), making their data free and easily accessible to researchers, ushering in the Data Golden Age (Tromble, 2021) of social media research. Scientists began using the data to study topics such as marketing, politics, psychology, crisis management, and computer science (Stieglitz et al., 2018; Liu, 2020). For the first time, scientists could examine the opinions, ideas, and emotions of hundreds of millions of people (Manovich, 2012). Leveraging big data offers novel perspectives across a wide range of inquiries, facilitating the study of social phenomena on a scale once deemed impossible (Tufekci, 2014). Examples of the wide variety of studies based on social media data include: a study at Cornell University using Twitter data to study seasonal mood changes across different cultures (Golder & Macy, 2011); a study at Instituto Superior Técnico using Twitter data to detect financial events that would likely impact stocks on the Dow Jones Average (Daniel et al., 2017); and a study at the University of Melbourne studying voter opinion and behavior during the 2013 Philippines national election (Pablo et al., 2014).

## Early Signs of Restriction: Privacy, Monetization, and API Access Limits

While the period between 2005-2010 was one of free and easily accessible social media data, beginning in 2010-2011, growing concerns about privacy, the increased commercial value of social media data, and the number of API requests outgrowing platforms' technological capacities spurred platforms to standardize and regulate access to their data (Boyd & Crawford, 2012; Bruns, 2019; Harris, 2010; Melanson, 2011). In 2011, following charges by the US Federal Trade Commission that they had changed privacy settings without notifying users, and misrepresented to users the amount of personal data that could be sold to 3rd party apps, Facebook agreed to make significant changes to its privacy policies and allow for independent privacy audits (Federal Trade Commission, 2011).

In 2011, Twitter forced the closure of the Twapperkeeper.com service (Bruns, 2019), which many researchers used to access data, and discontinued whitelisting, which allowed authorized API users to make a higher number of requests per hour (20,000 compared with 350 for non-whitelisted users) (Melanson, 2011). Following the decommissioning of whitelisting, researchers could pay the data reseller Gnip to continue their full access to data, or they could continue their free access by using either a throttled REST API (which limited users to 350 requests per hour), or the Streaming API, which limited requests to 1% of the total Twitter traffic (Felt, 2016; Chen et al., 2023). As a result, many researchers found themselves priced out of their previously available data source (Bruns, 2012). The relationship between social media platforms and researchers continued to decline following the 2014 ‘emotional contagion’ study in which Facebook collaborated with researchers at Cornell to study how emotions were spread between friends on Facebook, but was widely criticized for methodological flaws (Panger, 2016), and failing to obtain prior consent from the users involved in the study (“Editorial Expression of Concern,” 2014; Flick, 2016).

## The Cambridge Analytica Effect and Beyond

In 2018, the Cambridge Analytica scandal, in which Facebook data was used to create targeted political ads without user’s consent (Cadwalladr & Graham-Harrison, 2018), precipitated the “APIcalypse” after which social media companies dramatically limited access to their data for all purposes, including academic research (Tromble, 2021). However, while the Cambridge Analytica scandal was used as justification for subsequent changes to social media APIs (Tromble, 2021), the scandal was just the tipping point in a slowly widening gap between platforms and researchers (Bruns, 2019). According to scholars, platforms imposed unnecessarily sweeping restrictions on data access without accounting for the effects on research, raising concerns that these measures would undermine the valuable knowledge of social, cultural, economic, and political dynamics that influence everyday experiences (Tromble, 2021).

Following the scandal, Facebook substantially reduced the functionality of its Events, Groups, and Pages APIs (Schroepfer, 2018), which were the key Facebook APIs used by researchers to study public activity on the platform (Bruns, 2019). Twitter soon took its own action by requiring researchers to agree to a terms of use policy that included a list of prohibited subjects for study, including political beliefs, along with racial and religious topics (Alaimo, 2018). Subsequently, in 2019, Twitter began limiting developers and researchers to 100,000 requests per day (for their User and Mentions Timeline APIs), and requiring developers/researchers who needed more than 100k requests to go through a formal review process. The company stated this was to ensure greater scrutiny of those developers using the API the most, and to screen for inappropriate use of the data (Twitter, 2019). Some scholars argue that platforms exploited the scandal to curtail independent and critical scholarly examinations conducted in the public interest (Bruns, 2019). For example, in 2019, Facebook took explicit steps to disable a browser plugin used by the journalism project ProPublica to investigate political ads shown on Facebook (Bruns, 2019).

## Nominal Transparency and Selective Access through Platform-Led Research Initiatives

At the same time that platforms were limiting access to their APIs, several major platforms launched initiatives aimed at allaying calls for more data transparency. Facebook launched its Ad Library in 2018, which provided researchers with access to data related to advertisements shown on the platform (Bruns, 2019), as well as other funding opportunities for research on WhatsApp, Instagram, and Facebook, including its partnership with Social Science One, which intended to develop a scalable, privacy-preserving model for academic access to platform data (Tromble, 2021; DeGregorio et al., 2019). For its part, Twitter launched a call for research to help “measure the health of Twitter” (X, 2018), and subsequently awarded grants to two research teams (out of 230 applicants), with the intent to “increase the collective health, openness, and civility of the dialogue on our service” (Gadde & Gasca, 2018). Yet many researchers viewed these platform efforts as symbolic or strategically limited in scope, granting platforms control over what data, methods, and research questions were considered legitimate (Freelon, 2018; Walker et al., 2019).

## Reopening, Then Reversal: The Academic Research API and Musk-Era Privatization

In 2020, Twitter took a major step toward providing researchers more access to its data by releasing the Academic Research API v2, which allowed qualified researchers free and broad access to their data (Chen et al., 2023). This new API included access to the full Twitter archive (historical and streaming) with a monthly cap of 10 million tweets, the ability to access full conversations including original tweets along with all replies, and replies to replies, and up to 1000 filtering rules when downloading data (Tornes, 2021; Blakey, 2024). With the release of the Academic API, Twitter continued its legacy as a gold mine of data for researchers studying everything from the spread of misinformation, to social psychology and emergency management (Vosoughi et al., 2018; Golder & Macy, 2011; Yin et al., 2012).

This brief period of accessibility was abruptly reversed in 2023, following the purchase of Twitter by Elon Musk. Under the new ownership, Twitter began charging $42,000/month, or more depending on usage, for its Enterprise access level (X, 2024b), and now requires researchers to get permission each time they need to share tweet IDs with other researchers (X, 2024a), making peer review and research replicability much more difficult. Lacking sufficient funds to pay this monthly fee, academics are now scrambling to continue their research without this important data source (Ledford, 2023). This shift represents not only a retreat from data accessibility, but also a fundamental redefinition of who gets to observe and interpret public discourse, a dynamic increasingly governed by commercial platform interests, rather than scholarly or public imperatives (Balkin, 2018).

Scholars from across the research community argue that open access to social media data for research purposes is essential in order to continue important inquiries into human behavior and online safety. They argue that blocking access to APIs will require researchers to use much more time-intensive data collection methods, which will make it impossible for them to produce "large or representative samples of real-world events, such as social movements, elections, let alone state and non-state sponsored disinformation campaigns" (Walker et al., 2019). Maziyar Panahi, a researcher at the Institut des Systèmes Complex de Paris stated that "the shutdown [of Twitter's Academic API] had a profound impact on our work as researchers" (Blakey, 2024, p. 33). And Dr. Kristina Lerman, at the University of Southern California's Information Sciences Institute, stated that without data transparency, researchers will not be able to establish where online threats are coming from, and therefore cannot mitigate those risks (Blakey, 2024). As reported by the Columbia Journalism Review and Reuters, a survey by The Coalition for Independent Technology Research showed that over 100 research studies were impacted by the shutdown in 2023 alone (Gotfredsen, 2023; Dang, 2023).

## Data Governance, Platform Power, and the Future of Social Media Research

Governing bodies have started passing legislation, such as the European Union's Digital Services Act (DSA) of 2022 (Stiković, 2024), to require or incentivize technology companies (such as Twitter) to provide open access to their data. The DSA states that "very large online platforms" must grant researchers access to non-public data that "contributes to the detection, identification and understanding of systemic risks in the [European] Union" (Regulation 2022/2065). This mandate extends to researchers anywhere in the world, not just European Union nations, provided they meet requirements, such as affiliation with a non-profit institution, and their research addresses systemic risks in the European Union (Albert, 2024; European Commission, 2025). The United States, to this date, has not passed legislation granting researchers access to social media data, however, several federal bills have been proposed in recent years (Nonnecke & Carlton 2022; Vogus, 2022). These include: the Platform Accountability and Transparency Act of 2023 (Platform Accountability and Transparency Act, 2024b). This bill would have allowed researchers to submit proposals to the National Science Foundation, which could in turn compel platforms to disclose data. The Social Media Data Act of 2022 (Social Media DATA Act, 2022) would have required social media companies to provide researchers access to advertising data. Finally, the Digital Services Oversight and Safety Act of 2022 (Digital Services Oversight and Safety Act, 2022) would have provided researchers at the Federal Trade Commission access to data for the purpose of identifying illegal content on social media platforms. Although all three bills have been introduced in the U.S. Congress, each has stalled in committee without receiving a full chamber vote. The Platform Accountability and Transparency Act has been introduced twice, but legal scholars caution that it may face constitutional challenges and could be interpreted as a form of government censorship (Platform Accountability and Transparency Act, 2024a).

These developments reflect the longitudinal tensions that have been growing in the data governance landscape. As platforms have evolved from early, more open infrastructures of distributed communication such as email, Usenet, and independent forums (Abbate, 1999;

Gillespie, 2010; Plantin et al., 2018) to monolithic, for-profit platforms that act as active gatekeepers of social knowledge, they increasingly shape not only what users see, but also what researchers and the public can know (Venturini & Rogers, 2019; Verhulst & Young, 2022). Decisions about API access, pricing, and research restrictions are not merely technical or economic; they constitute exercises of platform power that structure the contours of academic inquiry and public accountability (van der Vlist et al., 2022). The 2023 pricing changes to Twitter's Enterprise API exemplify this shift, curtailing what had been a relatively accessible channel for studying mediated public discourse (Bruns & Weller, 2016; Freelon, 2018) and turning it into a prohibitively costly, selectively-gated service that limits scholarly access to critical digital information (Fiesler & Proferes, 2018; Tromble, 2021). Against this backdrop of existing work, this paper aims to highlight what may be lost if social media data continues to be cost prohibitive. To this end, using Twitter as a case study, we want to understand how many Twitter data-based academic papers, across the disciplines, were published during the years of available data (2006-2022), along with how many were published following the API's closure in 2023 to assess the decline in studies.

Prior studies have undertaken similar inquiries, but often with a much narrower, discipline-specific approach such as social media data used for predicting election results (Brito et al., 2021) and vaccine hesitancy (Cascini et al., 2022). Others have used the single search term "Twitter" across a limited number of databases, resulting in a broader, but less targeted and less complete corpus. For example, Karami et al. (2020) captured 18,849 studies and applied various computational analyses like frequency analysis, topic modeling, topic analysis, and trend exploration to quantify the ideas and concepts studied between 2006 and 2019. Yu & Munoz-Justicia (2020) conducted a bibliometric analysis of 19,205 Twitter-related studies, performing a performance analysis across five categories – Annual Scientific Production, Most Relevant Sources, Most Productive Authors, Most Cited Publications, and Most Relevant Keywords – along with a country-collaboration analysis and a thematic analysis to identify major research topics.

By synthesizing nearly two decades of scholarship across disciplines, our study fills a crucial empirical gap: the lack of a comprehensive understanding of how Twitter data has been used across academic fields, including during the year and a half since the API's closure. While prior reviews have focused on single topics or disciplines, or relied on broad keyword searches within a limited set of databases, our cross-disciplinary approach aims to reveal both the breadth and depth of Twitter data's role in scholarly research. And thus, we ask the following research questions:

RQ 1: How many academic studies have been published using Twitter data across the disciplines during the period when Twitter's free APIs were available (2006–2022), and how has this number changed since the Academic Research API's closure in 2023?

RQ 2: Which academic disciplines are most prominently represented in research utilizing Twitter data, and what are the common topics explored within these disciplines?

# Materials and Methods

## Data Collection

We sought to capture as many studies as possible that specifically utilized Twitter user data as their focus of inquiry. To identify these studies, we conducted a search across eight databases, and updated previous searches to include studies through December, 2024. Additionally, our study calculated incoming citations (papers that cited the papers in our dataset) for each study, along with the total incoming citations for the corpus, in order to quantify the influence of this body of work. Our study also identified prominent disciplines, topics, publication venues, date distributions, and illustrates the drop off in studies following the closure of Twitter's free API in early 2023. We collected a total of 33,306 unique studies, in 8914 distinct publication venues (journals and conferences), with 610,738 incoming citations, spanning 19 years, and across 16 disciplines. [1]

We began our search in 2006 since that was the year Twitter opened its API to academic researchers. Given the fractured landscape of literature databases, it was necessary to collect studies from a wide variety of sources in order to capture the maximum number of studies from every possible discipline. As a starting point, we conducted a broad search using Web of Science, one of the most comprehensive, multidisciplinary databases available. Searching the topics field, which included title, abstract, and author keywords (Topics = twitter NEAR/3 data OR twitter NEAR/3 api OR twitter NEAR/3 dataset), we located 4,826 articles. Utilizing Web of Science's built-in "Analyze Results" feature, we found that the top disciplines included: Computer Science, Engineering, Information Science, Communications, Public and Environmental Health, and Multidisciplinary Sciences. We then referenced the University of Washington's library guides for each of these disciplines to identify the most relevant research databases (see below), and then set about searching each database. All searches used some version of our initial search string, adjusting proximity operators as appropriate, and searching primarily in the topic, title, abstract, and keyword fields (Appendix 1). We also found that adding "NOT survey" to the string eliminated studies that simply used Twitter to disseminate surveys or find participants for data collection. Finally, we fine-tuned each search to include only journal articles, conference papers, dissertations, and preprints.

In Table 1 we list each database along with the number of results found, and the percentage of relevant studies within each results list. The statistical software, R, was used to randomize results for sampling. For each database (Appendix 1), a minimum of 50 sample studies were examined by hand to determine if they met one of three criteria: utilized Twitter data in the study, examined novel ways of extracting and studying Twitter data, or reviewed the literature of Twitter-based studies. To label sample studies relevant/not relevant, we found that most studies explicitly stated in their abstract if they utilized Twitter data. For example, “The researchers analyzed 100,000 tweets with hashtags #coronavirus . . . ” (Pandey et al., 2022, p. 1). In a minority of cases, when the abstracts were unclear, we examined methodology sections for confirmation. The most common reasons for labeling “not relevant” were studies that used Twitter to disseminate surveys, analyzed surveys about Twitter use, and studies that mentioned Twitter, but actually examined Sina Weibo, China’s Twitter alternative, or another social media platform.

---

[1] The 2024 version of this preprint contained an error in the reported citation count. It stated that 1,303,142 citations were identified; however, this figure was incorrect. When we originally used the Crossref API to retrieve citation counts, the API returned a default value of 999 for studies not found in the Crossref database. These values were erroneously included in the total citation count. The corrected citation count is **610,738**.

**Table 1**

| Database | Number of Results | Percent Relevant |
|---|---|---|
| Library and Information Science Source/Library Information Science and Technology Abstracts<br><br>LISS/LISTA (EBSCO databases) | 2,128 | 82% |
| Web of Science (SCI-EXPANDED, SSCI, AHCI, ESCI) | 13,427 | 82% |
| Global Health | 737 | 80% |
| ACM Digital Library | 2,664 | 92% |
| IEEE Xplore | 6,719 | 97% |
| Engineering Village (Compendex) | 24,941 | 86% |
| Engineering Village (Inspec) | 18,413 | 88% |

In the case of Engineering Village, the web-based database limits downloads to 1000 studies within a given search. With such a large number of relevant studies published in this database (over 43K), we deemed it essential to find another way to access these studies. Elsevier (publisher) offers two APIs for Engineering Village: the search API, and the retrieval API. We utilized R exclusively to access these data. Using the search string "(twitter AND data) OR (twitter AND api) OR (twitter AND dataset) NOT survey" we used the search API to obtain the "doc id" for each study, and then used the retrieval API to obtain metadata for a total of 43,354 studies. Extensive computational programming with R and Excel was required to unnest, clean, wrangle, and analyze the data. We combined the Engineering Village dataset with the dataset created from the other six databases, removed duplicates by DOI, title, and abstract, using R's "distinct" function, and randomized to create the final dataset. To quantify influence, we tabulated the incoming citations for each study in the dataset via the Crossref REST API. The final dataset (34,770 articles) includes: title, abstract, date of publication, manuscript date, document type, publisher (venue), publishing company, DOI, and citation count. Data and code available here.

## Data Analysis

The distribution of study dates was assembled in R (Graph 1). We ranked the top 100 publication venues, in R, by the number of published Twitter-based studies, then extracted the top 10 publication venues from this list and visualized the results in Graph 2. Next, we assigned disciplines by hand (Appendix 2) to each of the top 100 publication venues, calculated the percentage for each discipline, and visualized the results in Graph 3. To identify the most influential studies, and to provide a secondary analysis of disciplines within the corpus (Graph 4), we ranked the studies by their number of incoming citations, and then labeled the top 100 studies' disciplines by hand (Appendix 3).

Finally, to ascertain the main topics covered in the corpus, we manually analyzed the top cited studies from each discipline represented in the top 100 dataset (Appendix 3): Artificial Intelligence, Business, Computer Science, Data Science, Emergency Management, Information Science, Psychology, Public Health, Social Media, and Social Science. When possible, we selected the top 5 most cited studies from each discipline. However, several disciplines had fewer than 5 papers in the top 100 dataset: Psychology (2), Information Science (1), Emergency Management (3), so in those cases, all available studies were analyzed. We manually analyzed each title and abstract, noted the main themes, then grouped themes into common topics (Appendix 4).

## Results

For RQ 1—*What was the volume of studies published using Twitter data during the period of free access (2006–2022), and how did that volume change following the closure of the free API in 2023?*—we identified 26,817 studies published between 2006 and 2022. This period was characterized by a median annual growth rate of 25%. In 2023, following the closure of the free

API, 4,249 studies were published, representing a marginal increase of 0.6% compared to 2022. And in 2024, the number of studies declined to 3,712, a decrease of 13% from the previous year. Across the full period from 2006 to 2024, a total of 33,306 studies were identified (see Final Dataset). The first study to use Twitter data that we uncovered was "Why we twitter: Understanding microblogging usage and communities," published in August 2007 by Java et al. at the University of Maryland (Java et al., 2007). Examining the topological and geographical properties of Twitter's social network, the article explored the virtually uncharted territory of how people find each other and interact on social media. While Twitter's API became available in 2006, we did not find any articles published until this article in August, 2007. The studies we collected were published in 8914 distinct publication venues, with 610,738 incoming citations, over a span of 19 years, and across 16 broad disciplines. Graph 1 shows the spread of published studies between the years 2006 and 2024. Graph 2 shows the top 10 publication venues, ranked by the number of Twitter-based studies they each have published.

**Graph 1**

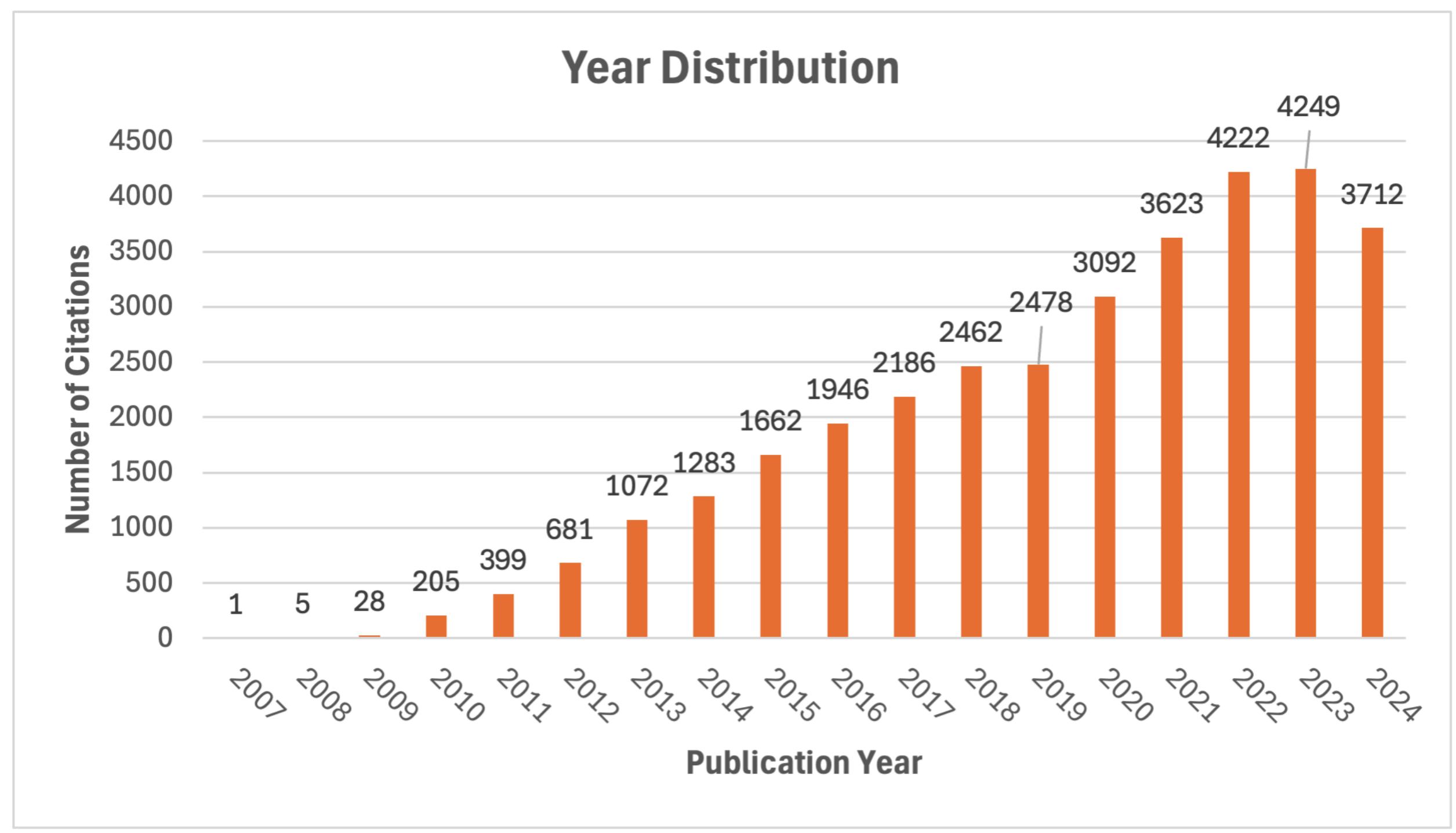

**Graph 2** (For full list of the top 100 publications, see Final_Dataset_and_ChartsDec2024, sheet Top Disciplines (by Venue) in Github repo):

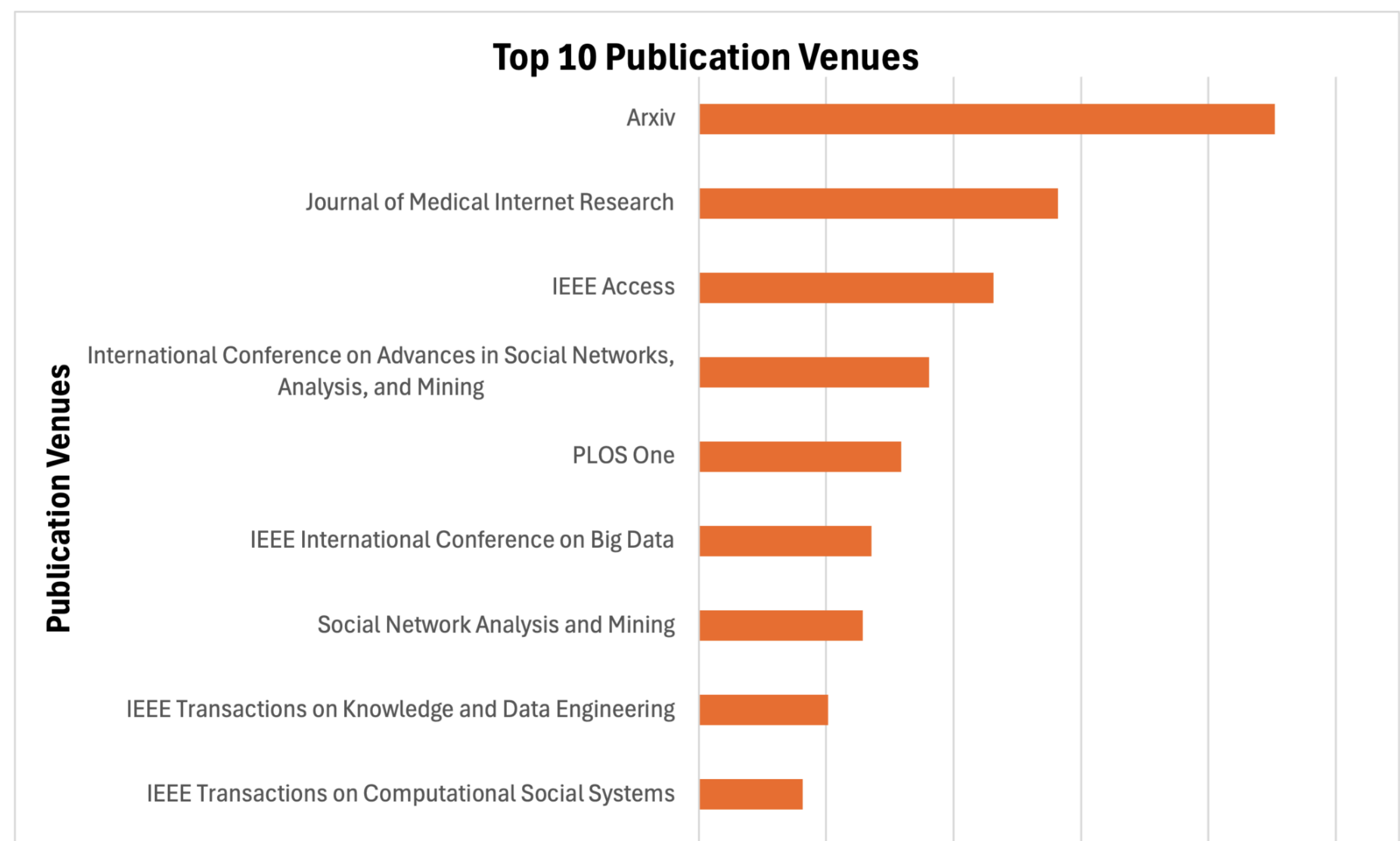

To address the first component of RQ 2—*Which academic disciplines are most prominently represented within the corpus?*—we employed a two-pronged methodology. First, Graph 3 shows the percentage of each discipline as determined by the top 100 publication venues' titles and/or website content (Appendix 2). Science/Engineering comprised 27% of the studies (percentages for all disciplines are rounded). Information Science comprised 14%, followed by Computer Science at 11%, and Data Science at 10%. Social Media comprised 9%, Medicine at 8%, Social Science and Internet Technology both had 7%, and the remaining ~8% were shared between Public Health, Artificial Intelligence, Human-Computer Interaction, Sustainability, and Emergency Response.

**Graph 3**

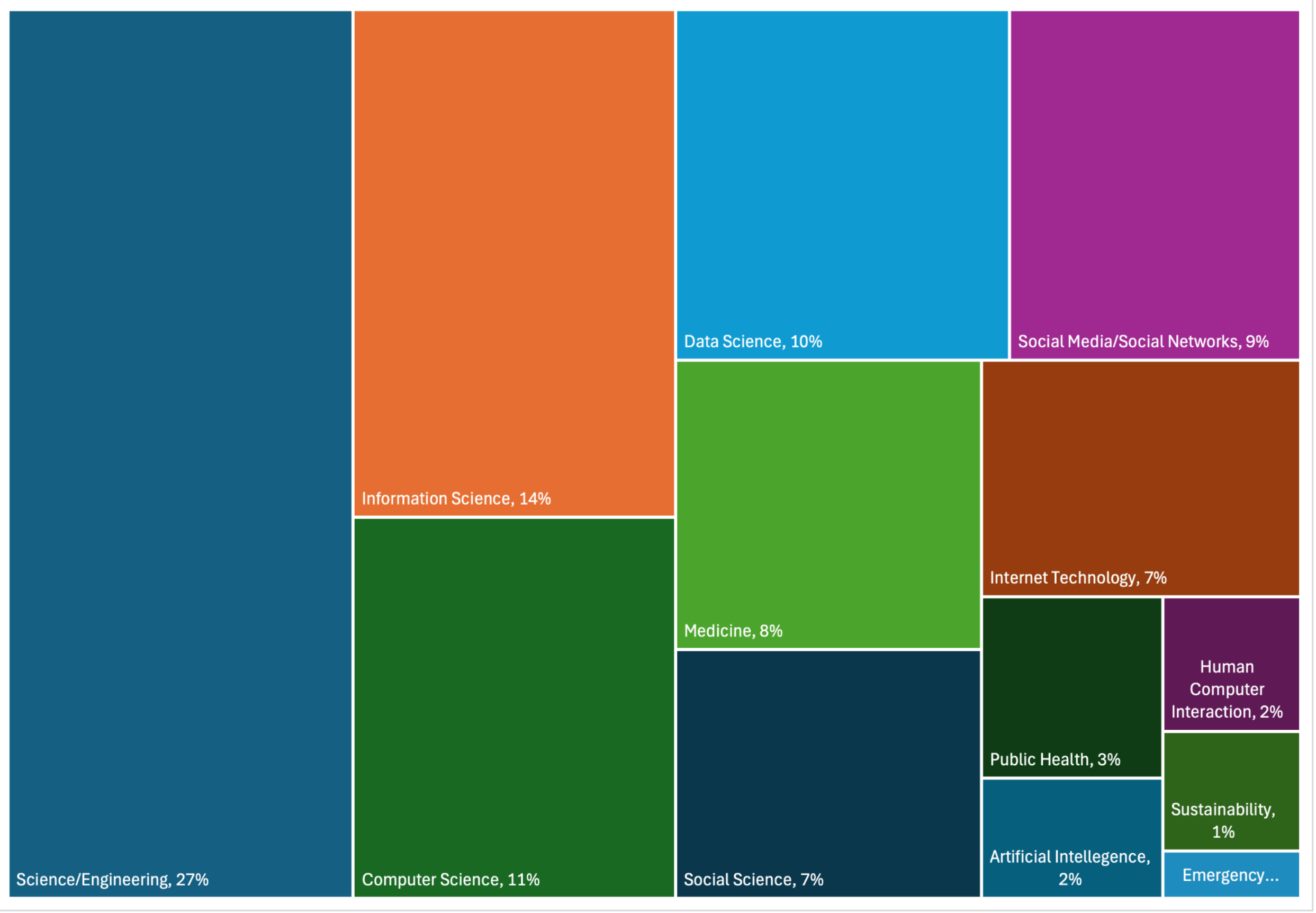

Second, Graph 4 shows the percentage of each discipline as determined by assigning disciplines to the top 100 most-cited studies (Appendix 3). In all, studies within the corpus were cited 610,738 times according to our Crossref analysis. It is important to note that this is only an approximation. There are many ways to count citations (ie. Crossref, Google Scholar, Web of Science), and each can differ considerably. Within these top-cited studies, Social Science comprised the highest number of studies at 24%, with a total of 19,113 citations. Data Science comprised 19% of the studies with a total of 15,033 citations, Social Media Studies comprised 14% with a total of 10,927 citations, and the remaining 43% included: Public Health, Computer Science, Artificial Intelligence, Business, Emergency Response, Information Science, Education, and Psychology.

**Graph 4**

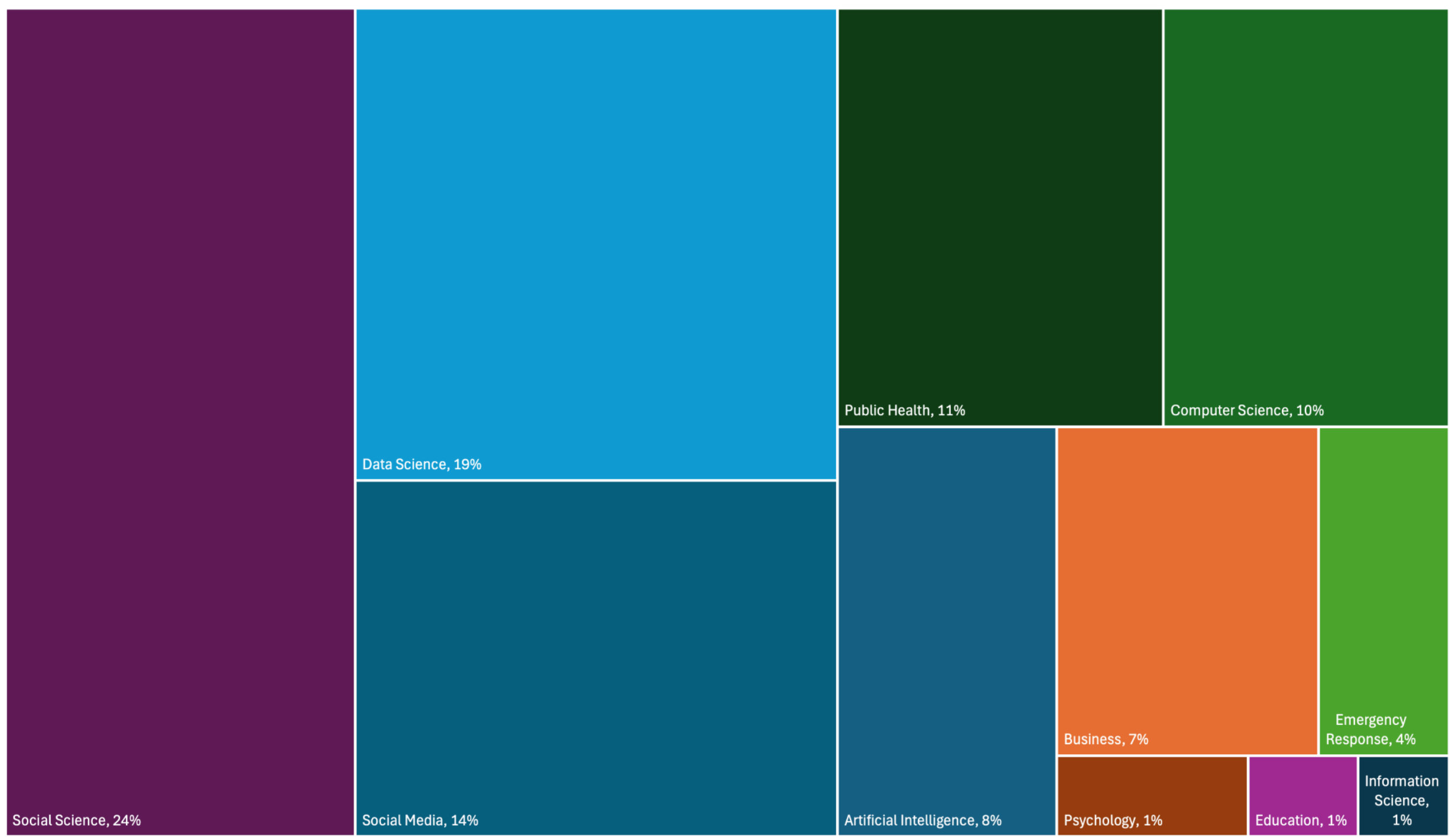

To address the second component of RQ 2—*What are the common topics explored within these disciplines?*—we present a summary in Table 2. The identified topics include: information dissemination, assessing the credibility of tweets, methodologies for conducting data research, detecting and analyzing major events, and studying human behavior (Appendix 4).

**Table 2**

| Main Topics | Examples: |
|---|---|
| Information Dissemination | • Dissemination factors and patterns<br>• How does political ideology impact dissemination?<br>• How does truth or credibility impact dissemination? |
| Information Integrity | • True vs fake news<br>• Disinformation<br>• Hate speech |
| Big Data Research Methodologies | • Sentiment analysis<br>• Topic modeling |
| | • Text Mining<br>• Text classification<br>• Neural networks<br>• Geolocating<br>• Data privacy<br>• Altmetrics |
| Detecting/Analyzing/Responding to Major Events | • Public health crises<br>• Emergency response and communication<br>• Combating rumors |
| Studying Human Behavior | • Political analysis<br>• Measuring public sentiment<br>• Mental health analysis<br>• Marketing/promotion of consumer products |

## Discussion

Between 2006 and 2022, Twitter data was widely used in academic research, increasing annually by a median rate of 25%, and spanning a wide swath of academic disciplines. However, the number of studies increased only marginally in 2023, and decreased by 13% in 2024 following the closure of its free API (Graph 1). We assume that much of the data used for studies published in 2024 were collected prior to Twitter's 2023 API shutdown, and thus the number of new studies are likely to continue declining in the coming years. We suggest a future study to collect these same data and analyze in a similar manner to obtain a more clear picture of the long term impact of Twitter's API shutdown.

In examining our first research question that asked how many studies used Twitter data during the period of free access (2006-2022), and how did usage change after the 2023 API closure, we identified 26,817 studies published between 2006 and 2022, with a median annual growth rate of 25%. After the API closure, output stagnated in 2023 and declined by 13% in 2024. This pattern reinforces the historical trajectory outlined in the literature review: as long as Twitter maintained relatively open data access, scholarly output expanded rapidly, but once access was abruptly curtailed, the research pipeline contracted in measurable ways. In other words, the publication record itself reflects the rise and fall of the Golden Age of open API research, and the subsequent retreat into a privatized, restricted model of platform data governance. In 2023 alone, following the API shutdown, over 100 research studies were cancelled or substantially impacted (Gotfredsen, 2023; Dang, 2023). One striking example was the closure of Botometer, a tool created at Indiana University that could detect if a Twitter account was a bot or a human and help fight the spread of disinformation (Gotfredsen, 2023).

In examining our second research question, which sought to understand which disciplines most prominently used Twitter data, and what major topics were explored, we identified 16 disciplines, including data science, internet technology, social science, psychology, public health, and emergency response. Common research themes included information dissemination, the spread of disinformation, big data methodologies, event detection, and human behavior. Notable studies examined the diffusion of true versus false news, finding that false, political news, spread faster and more widely than true news (Vosoughi et al., 2018); analyzed tweets as a form of word-of-mouth communication to inform corporate marketing strategies (Jansen et al., 2009); and tracked seasonal and daily mood changes and sleep patterns across multiple cultures (Golder & Macy, 2011).

Studies utilizing freely available Twitter data have had a broad impact, from tracking misinformation, to informing marketing campaigns, and helping governments identify emerging public health crises. By documenting and analyzing online behavior, these studies not only advance academic knowledge, they provide policymakers, business leaders, technologists, and non-profit and government agencies with the evidence needed to make informed decisions about platform governance, the development of new technologies, and public health and safety. The World Health Organization published a paper detailing how social media data is used by governments around the world to detect and track disease outbreaks, including the avian influenza (Fung & Wong, 2013). And the European Parliament's Subcommittee on Human Rights commissioned a study of disinformation's impact on democracies around the world (Colomina et al., 2021), in which multiple social media studies were cited, including Vosoughi et al. (2018).

In examining how changes in API access impact researchers' abilities to study online behavior and discourse, our findings show that restrictions significantly constrain research feasibility and scope. Scholars are forced to use resource-intensive methods that limit the scale and representativeness of their analyses. These constraints have clear applied implications: technologists should design data infrastructures that balance privacy with public-interest research needs; researchers must adapt by pooling resources, diversifying data sources, and advocating with policymakers for greater open access; and policymakers should consider legislative actions, such as the EU's Digital Services Act, to ensure researchers' have access to this vital data source. Ultimately, API governance is not a technical side note, but a structural determinant of what society can know about itself, making equitable, transparent access to social media data essential for informed scholarship and democratic accountability.

We employed two distinct methodologies to analyze the spread of disciplines within the corpus, resulting in two substantially different groupings. It is worth noting the difficulty we found in assigning disciplines. Several categories overlap (ie. Data Science, Computer Science, and Internet Technology), and our assignments were subjective. Another researcher might assign different disciplines, thus ending up with different percentage spreads in Graphs 3 and 4.

Our first methodology (publication venue-based) focused on the overall disciplines of the top 100 publication venues Graph 3. We analyzed the name of each top venue to determine the overall discipline (Appendix 2). For example, the International Journal Of Advanced Computer Science and Applications was assigned to the Computer Science discipline. In instances when the name alone was inconclusive, we explored the organization's website to confirm the discipline. For example, the journal name Multimedia Tools and Applications does not clearly state a discipline, thus further investigation into the journal's website was needed to reveal an overall discipline of Science/Engineering. We acknowledge that any one venue may contain a variety of disciplinary studies, therefore this strategy lacks specificity. At the same time, this strategy resulted in a wider array of disciplines than our study-based strategy, in particular the inclusion of Science/Engineering, and the differentiation between Data Science, Information Science, Computer Science, and Internet Technology. This may have been a result of the way publishers defined their venues.

Our second methodology (study-based) focused on the studies themselves to determine disciplines, rather than on the publication venue names (Graph 4). Disciplines were assigned by reading the title and abstract from each of the top 100 most-cited studies, and then choosing the discipline that fit best based on a set of definitions (Appendix 3). We believe this strategy provided a more accurate analysis of which disciplines were most strongly represented in the corpus. In these results (Graph 4), Social Science comprised 24% of the studies (up from 7% in the publication venue-based strategy). Perhaps this increase is a result of researchers publishing Social Science studies in non-Social Science publications, therefore leading us to label them under different disciplines in our first methodology. Similarly, Data Science comprised 19% of the studies (up from 10% in the publication-based strategy). We believe this difference may have occurred because Data Science studies were published in journals or conferences with overall disciplines of Computer Science or Internet Technology. Additionally, the study-based strategy illustrated the influence of this body of research in the academic world, with 610,738 incoming citations.

We wish to acknowledge three limitations to our study. While researchers use data from many different social media platforms (such as Facebook and Reddit), we chose to focus solely on studies utilizing Twitter data. Twitter has historically been the most common source of social media data due to its ease of use, open access, and its public nature, and thus offered the most comprehensive view into the topics and disciplines studied by researchers (Tromble, 2021). We also acknowledge that we did not search every available database. We aimed to search the largest and most comprehensive databases covering the widest variety of disciplines. Some databases were excluded (ie. Academic Search Complete and PubMed) because all results were duplicates from other databases, while another (Communication Source) was excluded because it yielded a relevancy rate well below 80%. We were able to collect 43,354 studies from Engineering Village via their API, however the many hours required to access these data could hamper future researchers with limited time. Additionally, we did not search the full-text of papers. Instead, we searched metadata including: title, abstract, keywords, publication venue, publication date, and others. While most of the databases we searched did not offer a full-text search option, two did (ACM Digital Library and IEEE Xplore), and a new search using the full-text field may produce additional papers. Future research could expand on the knowledge presented here, not only by updating the data to track the decline in Twitter data studies, but also by further examining the discipline specific impacts. Do some disciplines or topic areas have more funding and therefore the ability to continue accessing Twitter data even with the high price tag? Do other areas of study with less funding see a greater decline in published studies? What other differences exist across the disciplines as a result of the free API's closure?

## Conclusion

Since first released in 2006, Twitter's APIs have served as invaluable sources of data for researchers investigating diverse aspects of social life, including the spread of misinformation, dynamics of public discourse, crisis communication, and political behavior. However, in 2023, Twitter began charging $42,000/month for its Enterprise level. Lacking sufficient funds to pay this monthly fee, academics are now scrambling to continue their research. This paper illustrates the enormous value of social media data across the academic disciplines, and highlights what may be lost if social media data continues to be cost prohibitive. A search of 8 databases and 3 related APIs found that between 2006-2024, a total of 33,306 studies were published in 8,914 distinct publications venues, with 610,738 incoming citations, across 16 disciplines.

Our findings show that restrictions to data access significantly impair scholars' abilities to study some of the most pressing issues of our day. Policymakers should consider legislative actions, such as those in the EU's Digital Services Act of 2022 and the US's proposed Platform Accountability and Transparency Act of 2023, to ensure sustained access. Ultimately, API governance shapes what society is able to study and understand about itself, highlighting the critical importance of open access to social media data in academic research and public accountability. Alarmingly, while Twitter data studies have increased every year since 2006, following Twitter's decision to begin charging researchers for data in 2023, the number of studies decreased by 13% in 2024. This trend highlights a growing loss of empirical insight and access to real-time, public communication, raising concerns about the long-term consequences for studying society, technology, and global events in the era of big social media data.

**Appendix 1 (Data Collection: Databases/Search Strings/Sampling)**

*These data were originally collected and sampled in June, 2023. We recollected and updated our data in June, 2025, but used only results through December, 2024 for the final numbers seen in Table 1 and throughout the paper, however we continued to rely on the June, 2023 data for relevance percentages as seen below.

*Library and Information Science Source* and *Library Information Science and Technology Abstracts*
https://www.ebsco.com/products/research-databases/library-information-science-and-technology-abstracts.
“Twitter N3 data OR twitter N3 api OR twitter N3 dataset” NOT survey, *filtered for conferences, journals, and magazines only *All Text for all fields

- 1608 results
- 82% relevance (50 paper sample)

*Web of Science (SCI-EXPANDED, SSCI, AHCI, ESCI)*
https://clarivate.com/products/scientific-and-academic-research/research-discovery-and-workflow-solutions/webofscience-platform/
“(twitter AND data) OR (twitter AND api) OR (twitter AND dataset) NOT (survey)”

- 10811 results
- 82% relevance (75 paper sample)

*Global Health Database*
https://www.ebsco.com/products/research-databases/global-health
“(twitter AND data) OR (twitter AND api) OR (twitter AND dataset)”

- 536 results
- 80% relevance (50 paper sample)

*ACM Digital Library*
https://dl.acm.org/
“(twitter AND data) OR (twitter AND api) OR (twitter AND dataset) NOT (survey)” *abstracts only

- 1950 results
- 92% relevance (50 paper sample)

*IEEE Xplore*
https://ieeexplore.ieee.org/Xplore/home.jsp
“Twitter NEAR/3 data OR twitter NEAR/3 api OR twitter NEAR/3 dataset NOT survey” *filtered out books

- 3509 results
- 97% relevance (50 paper sample)

*Engineering Village API*

https://dev.elsevier.com/
Query = (((((twitter AND data) OR (twitter AND api) OR (twitter AND dataset) NOT survey) WN ALL)) NOT (({ch} OR {ip} OR {bk} OR {er} OR {tb} OR {ed}) WN DT))
*Compex Database*

- 20,813 results
- 86% relevance (50 paper sample)

*Inspec Database*

- 15,013 results
- 88% relevance (50 paper sample)

**Appendix 2 (Discipline definitions, and assignment of disciplines based on venues)**

*For detailed discipline assignments see Final_Dataset_and ChartsDec2024.xlsx at github.com/ryanmurt/Twitter (Top Disciplines (by Venue) '24 tab)
**Disciplines were assigned by examining the name of the journal or conference and/or the "About" section of the organization's website and applying the following definitions:
***Numbers based on data through December 2024 (collected in June 2025, but 2025 papers were removed since it was an incomplete year)

| Disciplines | Definitions used to assign |
|---|---|
| Social Science | Focus of study of human behavior, including psychology |
| Data Science | Focus is on the tools for manipulating and analyzing the data |
| Computer Science | Focus is on programing/designing the software tool |
| Social Media | Focus is on understanding how the social media tool works, how the technology functions. Similar to Social Science, but related to how human behavior is impacted by the technology, more than purely human tendencies in society. How the tool functions and how it enables social interaction. |
| Internet Technology | A broad category of studying all things internet/world wide web |
| Information Science | Focus of study is on locating, accessing or organizing information. |
| Science/Engineering | Focus of study is on a non-computer/information science field such as engineering (excluding physics). |
| Artificial Intelligence, Human-Computer Interaction, Physics, Public Health, Medicine, Sustainability, Emergency Response | These remaining disciplines are easily distinguished by name of publication and/or organization's website. |

**Appendix 3 (Discipline definitions, and assignment of disciplines based on study titles and abstracts)**

*For detailed discipline assignments see Final_Dataset_and_ChartsDec2024.xlsx at github.com/ryanmurt/Twitter (Full Citation Data (2024) tab)
**Disciplines were assigned by reading titles and abstracts and applying the following definitions:

| Disciplines: | Definitions used to assign |
|---|---|
| Social Science | A focus on social human behavior. How do people behave? This is often in the context of social media in this corpus, but always with a focus on human behavior. |
| Computer Science | Focus is on designing the software tool |
| Data Science | Focus is on using the tools to manipulate and analyze the data, ie. sentiment analysis, topic modeling, other tools for programming and data analysis. |
| Information Science | A focus on finding, organizing, and making data available |
| Social Media | A focus is on the technology itself, ie. studying or designing a tool used to identify influential users on Twitter. How does information spread on Twitter? Looks at how the technology works, rather than how people think and act. |
| Public Health | A focus on using Twitter to study public health issues, often looking at how a public health concern is discussed in tweets and retweets. |
| Business | A focus on business, marketing, promotions |
| Artificial Intelligence | A focus on the creation of AI or the uses of AI in social media and other technologies. |
| Psychology | A focus on human psychology within the context of Twitter and tweets. Often examining tweets to better understand a specific psychological question. |

| Education | A focus on the use of Twitter in educational settings. |
| --- | --- |
| Emergency Response | A focus on using Twitter to detect and respond to natural disasters, pandemics, and other emergencies. |

## Appendix 4 (Analysis of prominent topics)

*Taken from Final_Dataset_and_ChartsDec2024 at github.com/ryanmurt/Twitter, (Topics by Top 5 Papers) tab)

To ascertain the main topics covered in the corpus, we manually analyzed the top cited studies from each discipline represented in the top 100 dataset (Appendix 3): Artificial Intelligence, Business, Computer Science, Data Science, Emergency Management, Information Science, Psychology, Public Health, Social Media, and Social Science. When possible, the top 5 most cited studies from each discipline were analyzed. However, several disciplines had fewer than 5 papers in the top 100 dataset: Psychology (2), Information Science (1), Emergency Management (3), so in those cases, all available studies were analyzed.

### Topics taken from the most cited studies from each represented discipline:

**Social Science**

- Sentiment Analysis; Text Mining; Natural Language Processing
- Big Data as data source for studies; data ethics
- What is social media, how do people use it, and how do like-minded people connect with each other?
- Connection between emotion and sharing behavior; political communication
- Political ideology of Twitter users, and whether Twitter is just an echo chamber, or do people from different views interact. Yes, for political issues, people tend to just interact with like-minded users, however liberals were more likely to participate in cross-ideological interactions (on all topics).

**Social Media**

- Disinformation and information diffusion; Level of diffusion is greater for fake news than true news.
- Information diffusion - why/how people retweet
- What impact retweeting? Info diffusion patterns
- Geolocating users based on content of tweets
- Disinformation, social bots tend to spread tweets from disreputable sources and via users with high number of followers

**Public Health**

- Information dissemination, Covid, source credibility
- Content analysis about public health crisis; evaluation of Twitter as a data source to compliment surveys
- Health misinformation; most common topics and platforms in health disinfo? Solutions to the problem.
- Covid, public health, using social media to track and combat rumors

- Covid, topic analysis aimed at understanding and address public concerns/needs around public health crisis

**Psychology**
- Impact on effectuation processes when entrepreneurs use Twitter
- Measuring happiness

**Information Science**
- Meaning of altmetrics in scientific studies - do they correlate to actual citations?

**Emergency Management**
- Harnessing social media data for disaster relief coordination and decision making;
- Social media aids disaster response and damage assessment
- System for using Natural Language Processing to learn about disaster awareness during crises

**Data Science**
- Information credibility of news on Twitter
- Topic modeling as a methodology
- Big data methodologies and what the future will require of computational methods
- Big data methodologies, content, scope, challenges
- Analyzed how well anonymizing techniques work in social media data mining, as a way of protecting user privacy.

**Computer Science**
- Big data methodology: Bayesian
- Deep learning and its applications in Sentiment Analysis.
- Fake news detection using Event Adversarial Neural Network (EANN), which can derive event-invariant features (and thus be generalizable to many news stories)
- New methodology for sentiment analysis
- Twitter data processing in real-time

**Business**
- Brand sentiments on Twitter - comparing automated vs manual techniques for assessing how brands are talked about on Twitter
- Information diffusion. Attempting to predict what types of tweets will spread the fastest and furthest.
- brand-related user generated content and how they differ across different platforms
- measuring economic uncertainty with Twitter data and comparing to other measures.
- text mining to assess customer sentiment about brands and use that to inform marketing and sales decisions.

**Artificial Intelligence**

- a survey of text classification algorithms
- hate speech detection using deep learning architectures
- Twitter sentiment classification.
- Using neural networks to identify sentiment/opinions
- New deep learning methodology for sentiment analysis

**Common topics:**

- Sentiment Analysis, Text Mining, Topic Modeling
- Measuring Happiness
- Understanding the meaning of altmetrics in scientific studies
- Information Diffusion
- Disinformation: Detection, Diffusion Patterns, Combating Rumors, Source Credibility, News Credibility
- Hate Speech Detection
- Political Communication and Impact of Ideology
- Disaster Response and Analysis
- Geolocating Users
- Public Health: Covid, Disinformation, Tracking and addressing public concerns during crises
- Big Data computational methods: Neural Networks, Text Classification
- User Privacy
- Business: Brand sentiment, customer sentiment
- Measuring economic uncertainty

| **Main Topics Identified by Hand** | **Examples:** |
|---|---|
| Information dissemination | • Dissemination factors and patterns<br>• How does political ideology impact dissemination?<br>• How does truth or credibility impact dissemination? |
| Information integrity | • True vs fake news<br>• Disinformation<br>• Hate speech |
| Big Data research methodologies | • Sentiment analysis<br>• Topic modeling<br>• Text Mining<br>• Text classification |

| | • Neural networks<br>• Geolocating<br>• Data privacy<br>• Altmetrics |
|---|---|
| Detecting, analyzing, and responding to major events | • Public health crises<br>• Emergency response and communication<br>• Combating rumors |
| Studying human behavior | • Political analysis<br>• Measuring public sentiment<br>• Mental health analysis<br>• Marketing/promotion of consumer products |